\documentclass[twocolumn]{aastex62}

\usepackage{url}
\usepackage[section]{placeins}
\usepackage{graphicx,times}
\usepackage{amsmath} 
\usepackage{amssymb}

\newcommand{\molhy}{H$_2$}

\def\ltsima{$\; \buildrel < \over \sim \;$}
\def\simlt{\lower.5ex\hbox{\ltsima}}
\def\gtsima{$\; \buildrel > \over \sim \;$}
\def\simgt{\lower.5ex\hbox{\gtsima}}

\received{July 31, 2019}
\revised{October 2, 2019}
\accepted{October 15, 2019}

\submitjournal{ApJ Letters}

\shorttitle{Fuel for NGC~6240's Black Holes}
\shortauthors{Medling et al.}

\begin{document}

\title{How to Fuel an AGN: Mapping Circumnuclear Gas in NGC~6240 with ALMA}

\correspondingauthor{Anne M. Medling}
\email{anne.medling@utoledo.edu}

\author[0000-0001-7421-2944]{Anne M. Medling}
\altaffiliation{Hubble Fellow}
\affil{Ritter Astrophysical Research Center, University of Toledo, Toledo, OH 43606, USA}
\affil{Research School of Astronomy and Astrophysics, Australian National University, Canberra, ACT 2611, Australia}


\author[0000-0003-3474-1125]{George C. Privon}
\affil{Department of Astronomy, University of Florida, 211 Bryant Space Sciences Center, Gainesville, FL 32611, USA}

\author[0000-0003-0057-8892]{Loreto Barcos-Mu\~{n}oz}
\affil{National Radio Astronomy Observatory, 520 Edgemont Road, Charlottesville, VA 22903, USA}
\affil{Astronomy Department, University of Virginia, 530 McCormick Road, Charlottesville, VA 22904, USA}

\author[0000-0001-7568-6412]{Ezequiel Treister}
\affil{Instituto de Astrof\'{i}sica, Facultad de F\'{i}sica, Pontificia Universidad Cat\'{o}lica de Chile, Casilla 306, Santiago 22, Chile}

\author[0000-0003-0522-6941]{Claudia Cicone}
\altaffiliation{Marie Sk{\l}odowska-Curie Fellow}
\affil{INAF--Osservatorio Astronomico di Brera, Via Brera 28, I-20121 Milano, Italy}

\author[0000-0002-2985-7994]{Hugo Messias}
\affil{Joint ALMA Observatory and European Southern Observatory,
Alonso de C\'{o}rdova 3107, Casilla 19001, Vitacura, Santiago, Chile}

\author[0000-0002-1233-9998]{David B. Sanders}
\affil{Institute for Astronomy, 2680 Woodlawn Drive, University of Hawai'i, Honolulu, HI 96822, USA}

\author[0000-0002-0438-3323]{Nick Scoville}
\affil{California Institute of Technology, MC 249-17, 1200 East California Boulevard, Pasadena, CA 91125, USA}

\author[0000-0002-1912-0024]{Vivian U}
\affil{Department of Physics and Astronomy, 4129 Frederick Reines Hall, University of California, Irvine, CA 92697, USA}

\author[0000-0003-3498-2973]{Lee Armus}
\affil{Spitzer Science Center, California Institute of Technology, Pasadena, CA 91125, USA}

\author[0000-0002-8686-8737]{Franz E. Bauer}
\affil{Instituto de Astrof\'{i}sica, Facultad de F\'{i}sica, Pontificia Universidad Cat\'{o}lica de Chile, Casilla 306, Santiago 22, Chile}

\author[0000-0001-9910-3234]{Chin-Shin Chang}
\affil{Joint ALMA Observatory and European Southern Observatory,
Alonso de C\'{o}rdova 3107, Casilla 19001, Vitacura, Santiago, Chile}

\author[0000-0001-8627-4907]{Julia M. Comerford}
\affil{Department of Astrophysical and Planetary Sciences, University of Colorado, Boulder, CO 80309, USA }

\author[0000-0003-2638-1334]{Aaron S. Evans}
\affil{Astronomy Department, University of Virginia, 530 McCormick Road, Charlottesville, VA 22904, USA}
\affil{National Radio Astronomy Observatory, 520 Edgemont Road, Charlottesville, VA 22903, USA}

\author[0000-0003-0682-5436]{Claire E. Max}
\affil{Department of Astronomy \& Astrophysics, University of California, Santa Cruz, CA 95064, USA}

\author[0000-0002-2713-0628]{Francisco M{\"u}ller-S{\'a}nchez}
\affil{Department of Physics and Material Sciences, University of Memphis, Memphis, TN 38152, USA}

\author[0000-0001-6920-662X]{Neil Nagar}
\affil{Universidad de Concepci\'on, Departamento de Astronom\'ia, Casilla 160-C, Concepci\'on, Chile}

\author[0000-0002-5496-4118]{Kartik Sheth}
\affil{NASA Headquarters, 300 E Street SW, Washington, DC 20546, USA}





\begin{abstract}
Dynamical black hole mass measurements in some gas-rich galaxy mergers indicate that they are overmassive relative to their host galaxy properties.  Overmassive black holes in these systems present a conflict with the standard progression of galaxy merger - quasar evolution; an alternative explanation is that a nuclear concentration of molecular gas driven inward by the merger is affecting these dynamical black hole mass estimates.  We test for the presence of such gas near the two black holes in NGC~6240 using long-baseline ALMA Band 6 observations (beam size 0\farcs06 $\times$ 0\farcs03 or 30 pc$\times$15 pc).  We find (4.2-9.8) $\times10^{7}$ M$_\sun$ and (1.2-7.7) $\times10^{8}$ M$_{\sun}$ of molecular gas within the resolution limit of the original black hole mass measurements for the north and south black holes, respectively.   In the south nucleus, this measurement implies that 6-89\% of the original black hole mass measurement actually comes from molecular gas, resolving the tension in the original black hole scaling relations.  For the north, only 5\% to 11\% is coming from molecular gas, suggesting the north black hole is actually overmassive.  Our analysis provides the first measurement of significant molecular gas masses contaminating dynamical black hole mass measurements.  These high central molecular gas densities further present a challenge to theoretical black hole accretion prescriptions, which often assume accretion proceeds rapidly through the central 10 pc.
\end{abstract}



\section{Introduction}

Virtually every massive galaxy hosts a supermassive black hole at its center \citep[e.g.][]{Heckman14}.  
Using stellar and gas kinematics from adaptive optics assisted near-infrared integral field spectroscopy, 
\citet{Medling15_bh} measured dynamical masses of eleven black holes in nearby gas-rich major galaxy mergers and found the sample to be significantly overmassive relative to three black hole scaling relations: the M$_{\text{BH}}$ - $\sigma_{*}$ relation \citep{Ferrarese00,Gebhardt00,Tremaine02}, the M$_{\text{BH}}$ - L$_{\text{bulge}}$ relation \citep{Marconi03}, and the M$_{\text{BH}}$ - M$_{*}$ relation \citep{Kormendy95,Magorrian98}.  Finding overmassive black holes in ongoing mergers is surprising because we expect the optical quasar phase to happen at the end of a galaxy merger \citep[e.g.][]{Sanders88,Sanders96,Hopkins08}; thus, we expect that these black holes could still grow significantly in the next few hundred million years.  However, dynamical measurements only measure the total mass enclosed within the spatial resolution of the data; other mass from nuclear star clusters or gas must be subtracted to recover the true black hole mass.  In this Letter, we will use our ALMA observations to quantify how much gas mass is contaminating the M$_\text{BH}$ measurements of \citet{Medling15_bh}.

NGC~6240 \citep[16h52m58.9s +02d24m03s, z = 0.0243, log(L$_{\text{IR}}$/L$_{\sun})=11.93$; e.g.][]{Kim13} is a nearby gas-rich major merger hosting an active galactic nucleus (AGN) in both nuclei (``north'' and ``south''), separated by 735 pc \citep[1\farcs5;][]{Komossa03,Max07}.  Each nucleus hosts a small stellar disk \citep[effective radii 350$\pm$140 pc and 50$\pm$1 pc for north and south disks, respectively;][]{nucleardisks}.  Dynamical mass measurements of the black holes using the kinematics of these stellar disks found them overmassive by up to an order of magnitude relative to those expected from black hole scaling relations, placing the north black hole at $>$8.8$\times10^{8}$ M$_{\sun}$ and the south in the range (8.7-20) $\times10^{8}$ M$_{\sun}$ \citep{medling11,Medling15_bh}.  However, these measurements could be high if significant molecular gas surrounds the black holes, which would inflate the dynamical mass measurement.  Confirming these black hole masses is important for understanding how systems evolve along black hole scaling relations.

These long-baseline observations can resolve gas down to $\sim$15 pc scales, below the resolutions of the previous black hole mass measurements (FWHM $\sim$ 27 pc).  Using NGC~6240 as a test case, we will assess for the first time whether dynamical black hole mass measurements might be biased by the nuclear fuel reservoirs of the AGN themselves.

Throughout this work we adopt $H_0 = 70$\,km\,s$^{-1}$\,Mpc$^{-1}$,
$\Omega_{\rm m}$ = 0.28, and $\Omega_\Lambda$ = 0.72 \citep{Hinshaw09}.  The physical scale is thus 490 pc arcsec$^{-1}$, calculated using Ned Wright's Cosmology Calculator\footnote{Available at \url{http://www.astro.ucla.edu/~wright/CosmoCalc.html}.} \citep{Wright06}.


\section{Mapping CO(2-1) with ALMA \\
at High Spatial Resolution}
\label{obs}

We use long-baseline Band 6 imaging of CO(2-1) from ALMA, combining programs 2015.1.00370.S (PI: Treister) and 2015.1.00003.S \citep{Saito18}.  These observations from 2016 and 2017 span five different array configurations with baselines ranging from 15.1 to 14,969 m, for a total time on source of 44.5 minutes.  We reduced the data according to the standard ALMA data pipeline process\footnote{\url{https://almascience.nrao.edu/processing/science-pipeline}} with scaling according to measurements of ALMA's water vapor radiometers \citep{Maud17} and additional self-calibration steps.  The final spatial resolution of the dataset is 0\farcs06 $\times$ 0\farcs03 (30 pc $\times$ 15 pc).  Full details of the reduction can be found in Treister et al. (2019, submitted).
As described in Treister et al. (2019, submitted), the longest baseline observations have poorer phase stability than the others; poor phase stability means that not all flux might be recovered.  Removing these observations from the reduction reduces the resolution beyond what we need for the science goals described here, but we do so for comparison to estimate how much missing flux is affecting our high-resolution cube.  Compared to two different reductions (natural and Briggs-weighted with robust=1) of the lower-resolution cube, our high-resolution flux densities in the smallest apertures change by 3-20\%.  Because the fraction of flux we lose from the longest baseline observations depends on the spatial distribution of the gas itself, we apply no correction and simply note that, if anything, we expect our flux measurements to be low and thus our masses conservative.

\begin{figure*}
\centering
\includegraphics[scale=1.1]{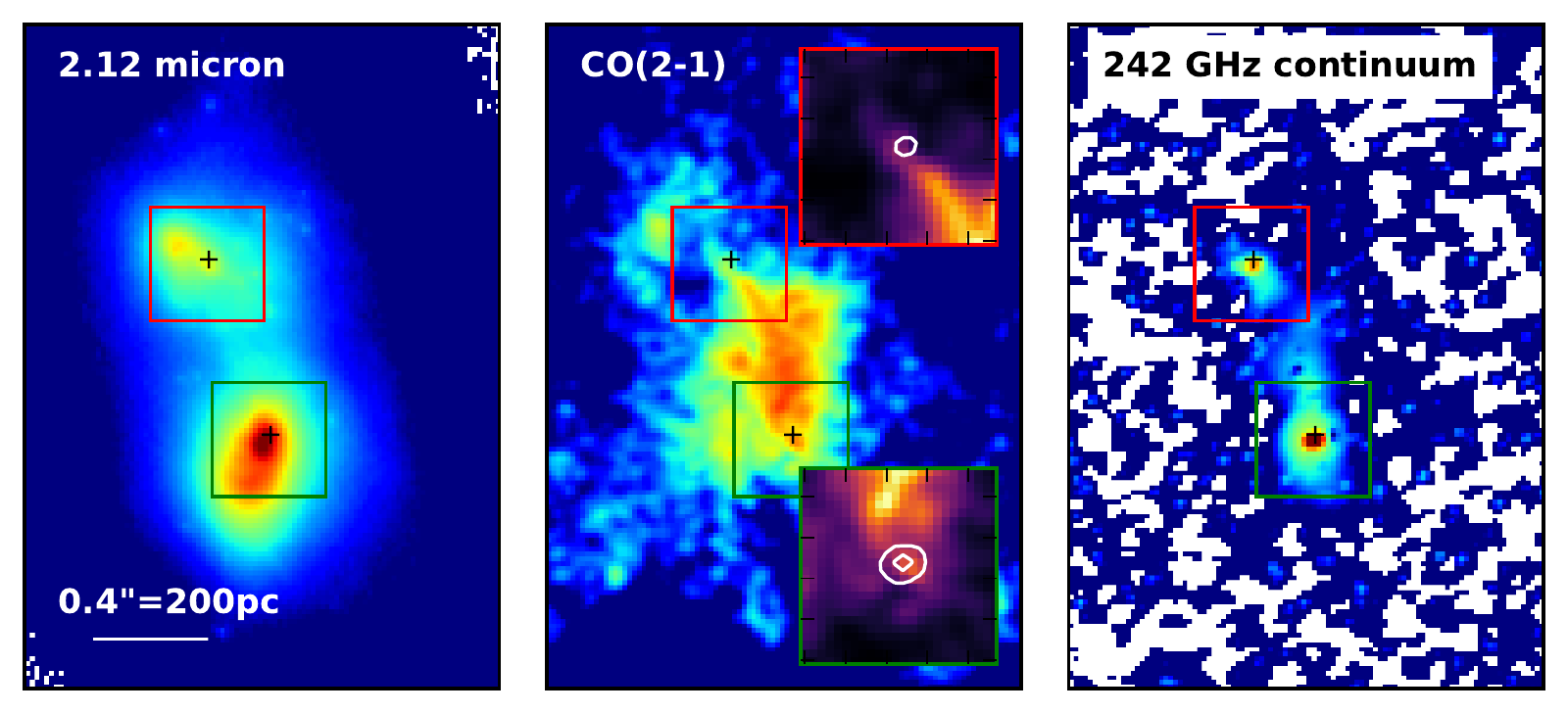}
\caption{\textbf{Left:} Keck NIRC2 K-band image of NGC~6240, highlighting the two nuclei, originally presented in \citet{Max05}.  \textbf{Center:} ALMA Band 6 moment 0 maps of CO(2-1) integrated over 1200 km s$^{-1}$.  Insets show nuclear regions in a different color scale for clarity, with continuum contours from right panel overlaid.  Images are rescaled in each panel to show structure; continuum contours are at 0.002 Jy beam$^{-1}$ in the north and (0.002, 0.02) Jy beam$^{-1}$ in the south, approximately 10$^3$ and 10$^4$ times the rms of the image.  \textbf{Right:} Rest frequency 242 GHz continuum contours from the same dataset peak at the locations of the two AGN.  Note that the millimeter continuum lines up with the kinematic centers of the K-band disks and not the photocenters, due to the large amount of dust present between the two nuclei that attenuates half of each disk even in the near-infrared.  Kinematic/continuum nuclear locations are consistent with those measured in X-ray and radio \citep[black crosses;][]{Max07}.  
}  
\label{COmap}
\end{figure*}

Figure~\ref{COmap} shows the CO(2-1) flux density map with 242 GHz restframe continuum contours overlaid in white identifying the locations of the two AGN.  The continuum map shows two peaks aligned with the locations of the AGN \citep[measured from X-ray and radio;][]{Max07} that are resolved into extended structures.  The CO(2-1) emission, on the other hand, follows a clumpy ribbon between the nuclei and gives little clue to the locations of the nuclei.  This ribbon is also detected in continuum, but it is much fainter than the nuclei.  The complex structure of the CO data are analyzed in depth in Treister et al. (2019, submitted); here we focus only on the flux and mass within the black holes' spheres of influence.  

\section{Results: How Much of an AGN's Measured Mass is Molecular Gas?}
\label{discussion}

\subsection{Previous Insights into Central Gas Masses}

Many hydrodynamic simulations of galaxies use subgrid prescriptions based on Bondi-Hoyle accretion \citep{Bondi44, Bondi52, Springel05_feedback}, which relies on the assumption that gas accreting onto the black hole has no net angular momentum.  On the contrary, high-resolution simulations \citep{Mayer07} and observations \citep[e.g.][]{Downes98, Bryant99, Sakamoto99, nucleardisks} both show that gas-rich systems form gaseous nuclear disks on scales of a few tens of parsecs, reaching scales at least down to the torus scale \citep[a few parsecs; e.g.][]{Gallimore16, GarciaBurillo16, Davis17,Davis18,Imanishi18,Izumi18, AlonsoHerrero18, AlonsoHerrero19,Combes19}.  Gas accreting onto the black hole must therefore dissipate its angular momentum in the process.  Viscosity in these parsec-scale gas disks is likely the dominant mechanism for removing angular momentum, and can delay accretion of most of the gas by a viscous timescale \citep{Power11,Wurster13}, which in supermassive black hole systems can reach up to a Hubble time \citep{King08}.  

Three-dimensional radiation-hydrodynamic simulations of AGN-driven fountains estimate that the cold gas ($<$40 K) within r$<$16 pc totals about 50\% of the black hole mass \citep{Wada12, Wada16}.  Additionally, recent long-baseline ALMA observations of Arp~220's west nucleus show that about half of the dynamical mass (7 $\times 10^8$ out of 1.5 $\times 10^9$ M$_{\sun}$) within 70 pc can be attributed to molecular gas \citep{Scoville17}.  \citet{Combes19} present a recent suite of ALMA observations of Seyfert nuclei on 6-27 pc scales, finding that molecular mass can also account for the majority of their dynamical masses at these radii.  However, we note that dynamical black hole mass measurements in several early-type galaxies using ALMA found negligible contributions from molecular gas in their nuclear regions \citep[e.g.][]{Barth16,Davis18,Boizelle19} 

\subsection{Measuring the Central Gas Mass Profiles in NGC~6240}

Because \molhy~lacks a permanent dipole moment, its pure rotational transitions are forbidden; asymmetric proxy molecules like CO are often used instead, but their use can be contentious.  We therefore measure the mass in two independent ways: by using an $\alpha_{\text{CO}}$ conversion from the CO(2-1) emission and by converting the dust continuum to a gas mass by assuming a dust-to-gas ratio, described in the following paragraphs.  

\begin{figure*}
\centering
\includegraphics[scale=.85]{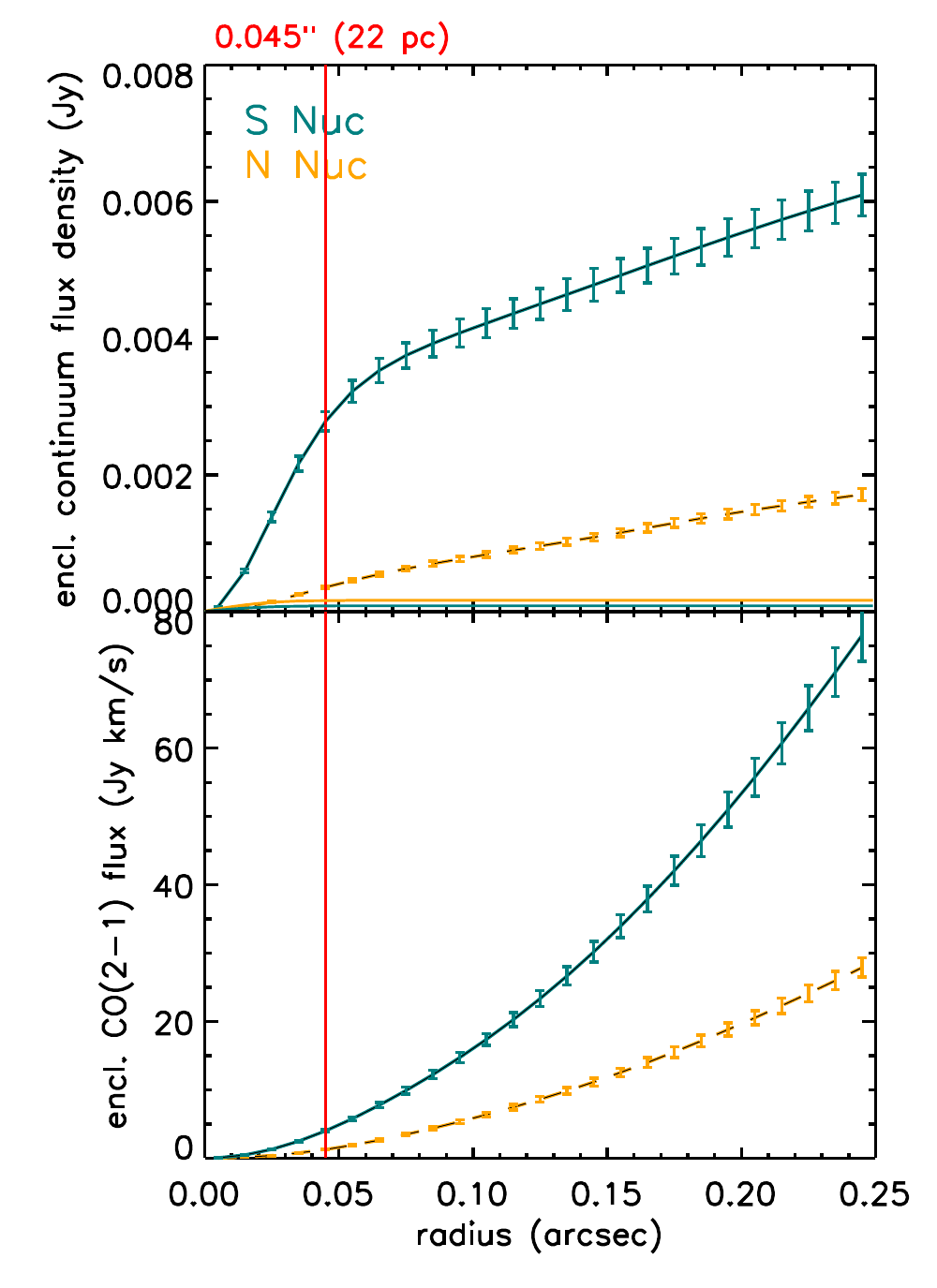}
\includegraphics[scale=.85]{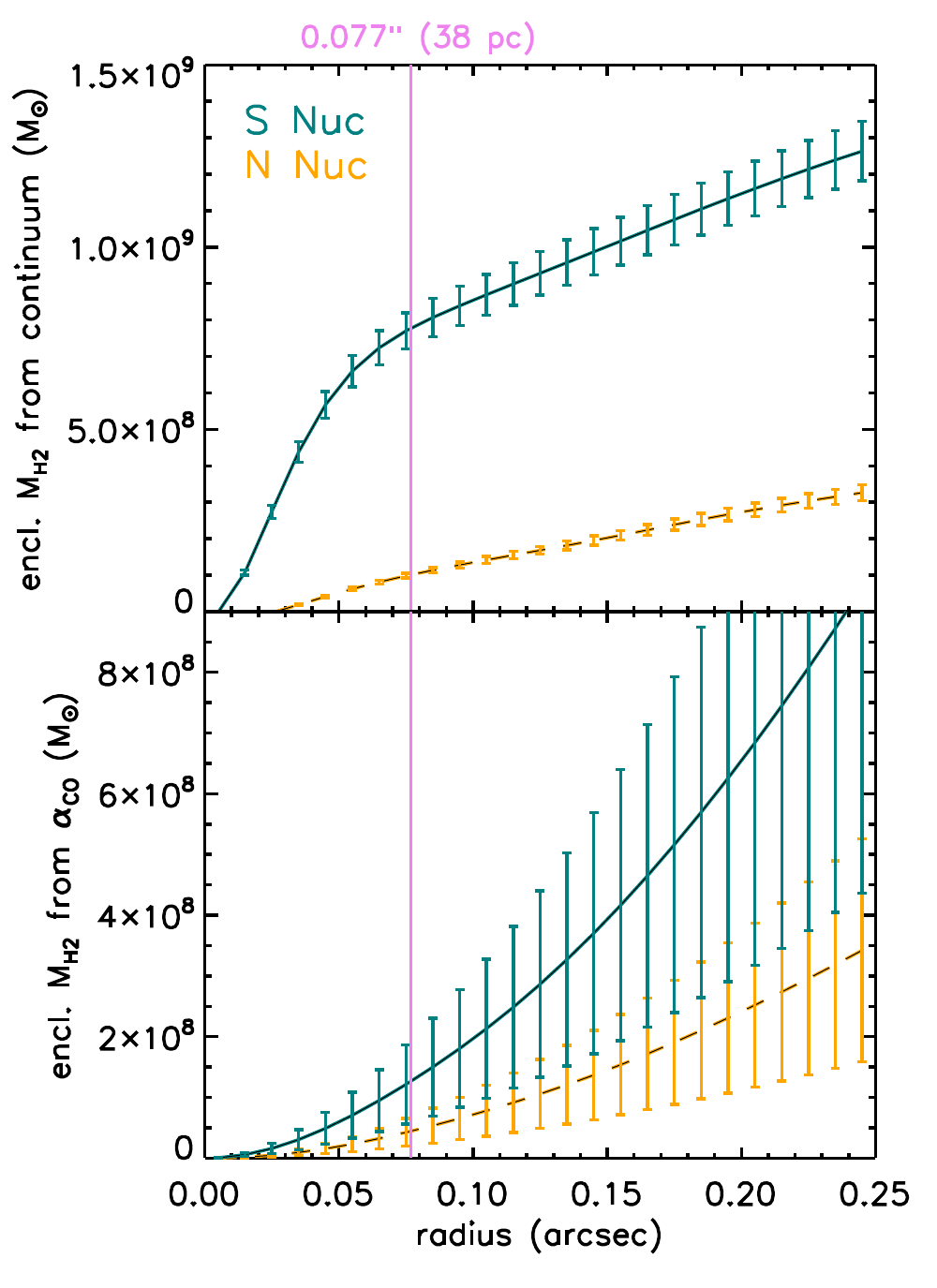}
\caption{Left: Enclosed flux profiles for the north (N Nuc, orange) and south (S Nuc, teal) nuclei in the continuum (top) and CO(2-1) emission (bottom), as a function of aperture size.  Nearly horizontal lines show the predicted enclosed synchrotron contribution based on long-wavelength radio measurements of the AGN, and are subtracted from fluxes before mass conversion.  Right: Enclosed \molhy~mass profiles using continuum conversion (top) and CO(2-1) emission (bottom), as described in the text.  The red vertical line in the left panel shows the angular resolution of the ALMA beam (0\farcs06 $\times$ 0\farcs03); the purple line in the right panel shows the angular resolution limit of the previously measured black hole masses in NGC~6240 \citep[0\farcs077 = 38pc;][]{Medling15_bh}.  Error bars in the left panel come from the rms of the ALMA cubes, added in quadrature to a 5\% absolute flux calibration error.  Errors in the top right panel include both flux errors and statistical errors from the \citet{Scoville16} calibration, but not errors in the temperature scaling.  The large error bars in the bottom right panel are due to the corresponding error in $\alpha_{\text{CO}}$.  
}  
\label{massprofiles}
\end{figure*}

To measure the mass from the CO(2-1) emission, we use the $\alpha_{\text{CO}}$ conversion factor calibrated for NGC~6240 by \citet{cicone18} to translate L$^\prime_{\text{CO}}$ to total molecular mass, including both \molhy~and helium.  They use spatially resolved measurements of [C I] $^{3}P_{1}$--$^{3}P_{0}$ and CO(1-0) taken with ALMA to calculate $\alpha_{\text{CO}}$ directly in a spatially varying manner, and for systemic vs outflowing gas; they incorporate the CO(2-1) observations described here to provide a spatially resolved calculation of $r_{21} \equiv L^{\prime}_{\text{CO(2-1)}} / L^{\prime}_{\text{CO(1-0)}}$.  We use the luminosities measured in their central box (2\arcsec $\times$ 2\arcsec, which contains both black holes), which produce $\alpha_{\text{CO(1-0)}} = 2.3 \pm 1.2$ M$_{\sun}$ (K km s$^{-1}$ pc$^{2}$)$^{-1}$ and $r_{21} = 1.26 \pm 0.14$, resulting in a final $\alpha_{\text{CO(2-1)}} = 1.83 \pm 0.97$ M$_{\sun}$ (K km s$^{-1}$ pc$^{2}$)$^{-1}$.  We note these quantities refer to the total gas reservoir within the central box, including both outflow and non-outflow components, which are heavily blended in these central resolutions at the spatial resolution of the \citet{cicone18} data.

Continuum emission in the submillimeter regime probes the Rayleigh-Jeans tail of the far-infrared dust emission.  We use the calibrated relation between continuum flux density and \molhy~mass of \citet{Scoville16}\footnote{Using measurements of L$_{850\mu\text{m}}$ from \citet{Klaas01} and M$_\text{mol}$ from \citet{cicone18}, we find that NGC~6240 globally has a calibration about 2$\sigma$ above that of \citet{Scoville16}.  However, the majority of the ISM in NGC~6240, between the nuclei, is undergoing a strong shock that likely affects the global ratio strongly \citep[e.g.][]{Meijerink13}.  We suspect that the nuclei are more typical and therefore better represented by the \citet{Scoville16} calibration.}.  This calibration has updated their earlier relations \citep{Scoville15} to include helium mass in the \molhy~mass and a larger sample of test galaxies.  However, because the global relation of \citet{Scoville16} assumes a dust temperature of 25 K, we scale the relation linearly to a mass-weighted dust temperature of 100 K, following the assumption made in Arp~220's nuclear regions on similar scales \citep{Scoville15}.  If the mean-weighted dust temperature is higher, the corresponding implied molecular gas mass will be lower by the same factor.   In the subsequent calculations, we include the statistical errors from the \citet{Scoville16} conversion, but do not include an extra term to account for errors in the dust temperature, beyond the aforementioned scaling.

Synchrotron emission from the AGN could also be contributing meaningfully to the continuum flux at such small spatial scales.  We estimate and subtract the synchrotron emission by scaling the MERLIN 5 GHz central beam flux from \citet{Beswick01} using a spectral index of $\nu^{-1.06}$ based on peak emission obtained from the Very Large Array (VLA) at 8.4 GHz \citep{Carral90} and 15 GHz \citep{Colbert94}.  Note that VLBA observations at 2.4 GHz and 8.4 GHz \citep{Gallimore04} predict an upper limit for the spectral index between these frequencies of -1.8 and -0.2 for the north and south nuclei, respectively, based on nondetections at 8.4 GHz.  However, the angular resolution of our ALMA observations include the point sources detected with VLBA plus some extended emission detected by MERLIN and the VLA; a matched resolution observation of the continuum at lower frequencies is needed to more precisely measure the synchrotron contribution at 242 GHz (e.g. with long-baseline ALMA band 3 and ngVLA observations).  Using these best-available observations,
the contributions here to our 242 GHz continuum are $1.6\times10^{-4}$ Jy and $8.2\times10^{-5}$ Jy for the north and south nuclei, respectively.  These fluxes are subtracted from our measured continuum fluxes before converting to gas mass.

In Figure~\ref{massprofiles} we show the enclosed continuum and CO(2-1) flux profiles and the corresponding derived enclosed mass profiles.  The south nucleus is considerably brighter in both CO(2-1) and continuum emission than the north nucleus.  The continuum flux also shows a more definitive knee in the radial profile, suggesting that it traces a distinct nuclear mass of gas.  The CO(2-1) emission might instead be contaminated by 
the gas bridge that connects the nuclei, where the bulk of the \molhy~mass resides (Treister et al. 2019, submitted).
The difference in morphology between CO(2-1) and continuum demonstrates the major uncertainty in measuring \molhy~mass.  That is, the assumptions that went into each of our mass calculations somewhere break down, and NGC~6240 might exhibit a spatially varying $\alpha_\text{CO}$, dust temperature, and/or dust-to-gas ratio.  However, these two mass estimates only differ by a factor of a few, so we propagate both estimates through our calculations and use them as an indication of our systematic error in the correction.  We further note that we estimate the molecular gas mass present on these scales rather than the total gas mass; i.e. we are assuming that the molecular gas phase and the black hole account for all of our previously measured dynamical masses.

We further note an additional uncertainty: we are measuring circumnuclear flux \textit{in projection} and assuming that it corresponds to all the circumnuclear mass; in truth, these fluxes are upper limits to the amount of flux emitted by gas in the nuclei themselves.  The continuum emission is spatially concentrated on the nuclei and therefore is probably truly circumnuclear rather than a projection effect.  However, the CO(2-1) emission map (Figure~\ref{COmap}) is much more complex.  As discussed in Treister et al. (2019, submitted), the kinematics of the CO in the nuclei do not appear to match the kinematics of the stellar nuclear disks from \citet{Medling15_bh}, which would have strongly suggested a nuclear nature.  Instead, we consider two possibilities: either the CO-emitting gas is still circumnuclear but has non-Keplerian kinematics (perhaps in the process of being stirred up / ejected by the AGN), or some/all of the CO-emitting gas is merely in projection and should not be included in our circumnuclear gas mass measurements.  Therefore, one might consider the CO-implied mass measurement an upper limit.  Because the continuum-implied mass profiles are higher in both nuclei, the caveats around CO should not change our conclusions.

\subsection{Interpreting the Central Gas Masses}

The original OSIRIS adaptive optics integral field spectroscopy used to measure the black hole masses by \citet{Medling15_bh} had a plate scale of 0\farcs035 pixel$^{-1}$.  The best resolution achievable by this plate scale is 0\farcs077 (38 pc), a Nyquist sampling of 2.2 pixels.  We therefore set 38 pc as our aperture in Figure~\ref{massprofiles} to calculate the gas masses contaminating the original OSIRIS measurements (Table~\ref{tbl:numbers}).  We find $9.8 \pm 0.7 \times10^{7}$ M$_{\sun}$ of gas (from the continuum emission) and $4.2 \pm 2.3 \times10^{7}$ M$_{\sun}$ of gas (from the CO emission) for the north nucleus and $7.7 \pm 0.5 \times10^{8}$ M$_{\sun}$ of gas (from the continuum emission) and $1.2 \pm 0.7 \times10^{8}$ M$_{\sun}$ of gas (from the CO emission) for the south nucleus.  This gas mass implies that 5\%-11\% (6\%-89\%) of the original dynamical mass measurement of the north (south) black hole is actually contamination by gas.

\begin{figure*}
\centering
\includegraphics[scale=.94, trim=.5cm 0cm 0cm 0cm, clip]{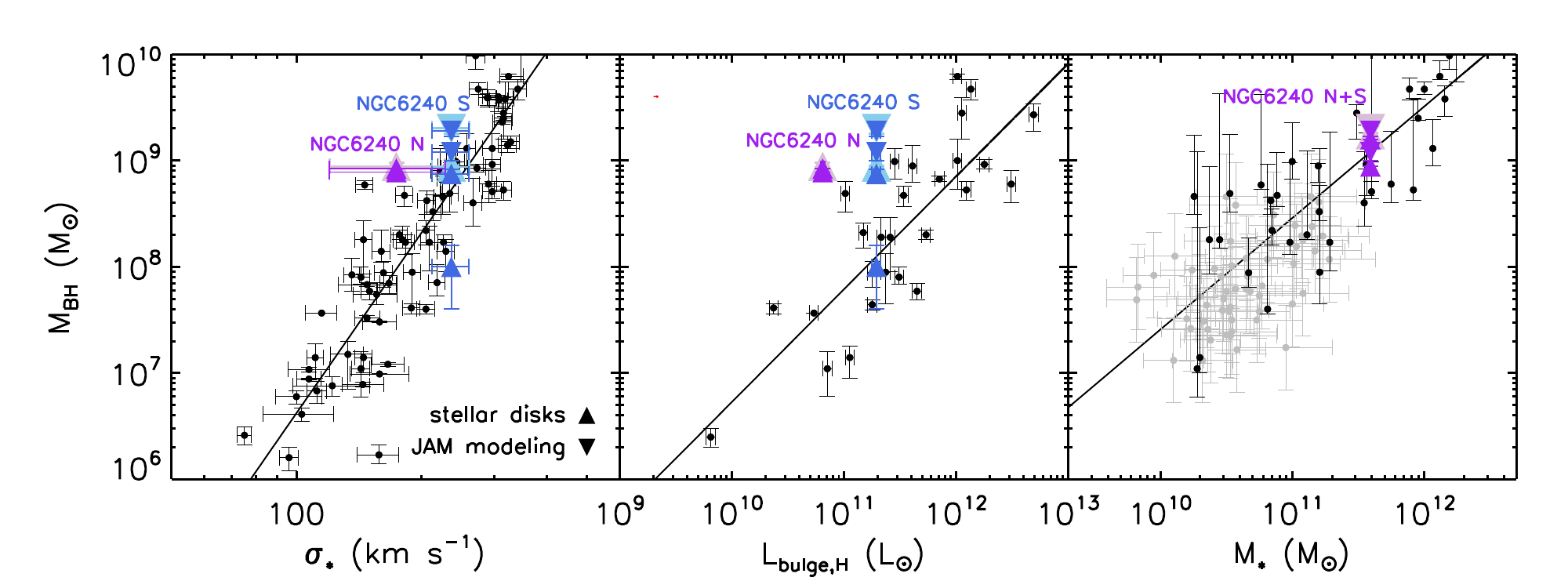}
\caption{Black hole scaling relations -- black hole mass M$_{\text{BH}}$ vs $\sigma_{*}$ (left), bulge luminosity L$_{\text{bulge}}$ (center) and total stellar mass M$_{*}$ (right) -- as in \citet{Medling15_bh}, showing original NGC~6240 black hole mass measurements (light purple for north, light blue for south) and new corrected black hole mass measurements in darker corresponding colors.  In the total stellar mass plot, the two black hole masses are summed and shown in purple.  Gas mass corrections to these measurements are significant in the south nucleus and represent a substantial reservoir of molecular gas in the central $\sim$40 pc, waiting to either accrete onto the black hole or be ejected by future molecular outflows.  The large mass of gas within the sphere of influence of the south black hole caused it to appear overmassive, when in fact its mass is consistent with black hole scaling relations.  The north black hole does not have much corresponding gas and still appears on the upper edge of the scatter of scaling relations, although the sum of the black holes is still consistent with the implication from the total stellar mass.
}  
\label{scalingrelations}
\end{figure*}

By correcting for the gas mass, the new black hole masses fall closer to traditional black hole scaling relations (Figure~\ref{scalingrelations}).  The large correction to the south black hole's mass (blue in Figure~\ref{scalingrelations}) draws its measurement comfortably into the scatter of scaling relations for quiescent elliptical galaxies.  The correction to the north black hole's mass is comparatively minor, and is insufficient to resolve its tension.  However, we note that the sum of the north and south black holes' corrected masses is still consistent with that expected from the stellar mass of the combined system -- that is, if there were no new star formation or black hole growth during the rest of the merger, the final system would lie on the M$_\text{BH}$ - M$_{*}$ relation.  The remaining tension in the north black hole's mass in Figure~\ref{scalingrelations}a,b may therefore instead be due to the unsettled nature of the corresponding bulge while the merger is ongoing.  More spatially resolved submillimeter observations of similar black holes are needed in order to determine which scenario is more typical.

\begin{table*}
\begin{centering}
\caption{Observed and Derived Quantities from the North and South Nuclei}\label{tbl:numbers}
\begin{tabular}{|l|c|c|c|}
\hline
& North Nucleus (Disk Model) & South Nucleus (Disk Model) & South Nucleus (JAM Model)\\ 
\hline
Dynamical M$_{\text{central}}$ (M$_{\sun}$)\tablenotemark{a} &  $> 8.8\substack{+0.7 \\-0.1 } \times10^8$  & $>8.7 \pm 0.3 \times 10^8$ & $<2.0 \pm 0.2 \times 10^9$ \\
\hline
Continuum flux (Jy)\tablenotemark{b}  & $6.3\pm0.3\times10^{-4}$ & \multicolumn{2} {c|} {$3.7\pm0.2\times10^{-3}$ }\\
Synchrotron contribution (Jy)\tablenotemark{c} & $1.6 \times 10^{-4}$ & \multicolumn{2} {c|} {$8.2 \times 10^{-5}$ } \\
M$_{\text{H}_{2},\text{continuum}}$   (M$_{\sun}$)\tablenotemark{b}   & $9.8 \pm 0.7 \times 10^7$ & \multicolumn{2} {c|} {$7.7\pm 0.5\times10^8$} \\
M$_{\text{BH,corrected,cont}}$ (M$_{\sun}$)\tablenotemark{d}  & $> 7.8\substack{+1.0 \\-0.7 } \times10^8$& $>1.0 \pm 0.6 \times 10^8$ &  $<1.2 \pm 0.2 \times 10^9$ \\
M$_{\text{BH}}$ Correction &11.1\% & 88.5\% & 38.5\%\\
\hline
CO(2-1) flux (Jy km s$^{-1}$)\tablenotemark{b}  & $3.5\pm0.2$ & \multicolumn{2} {c|} {$9.9\pm0.5$} \\
M$_{\text{H}_{2},\text{CO}}$  (M$_{\sun}$)\tablenotemark{b}   & $4.2\pm2.3\times10^7 $ & \multicolumn{2} {c|} {$1.2\pm0.7\times10^8$}\\
M$_{\text{BH,corrected,CO}}$ (M$_{\sun}$)\tablenotemark{d}  & $> 8.4\substack{+0.7 \\-0.1 }\times10^8$ & $>7.5 \pm 0.8 \times 10^8$ & $<1.9 \pm 0.2 \times 10^9$ \\
M$_{\text{BH}}$ Correction & 4.8\%& 14.0\% & 6.1\%\\

\hline
\end{tabular}
\end{centering}
\tablenotetext{a}{Central dynamical masses were reported as black hole masses in \citet{Medling15_bh} using Keplerian disk models and/or Jeans Axisymmetric Mass models.}
\tablenotetext{b}{Enclosed fluxes and gas masses are measured using an aperture of radius 0\farcs077 (38 pc), the resolution limit of the original dynamical mass measurements in \citet{Medling15_bh}.  In both cases, the molecular mass includes He.  The continuum flux is converted to gas mass following the relation of \citet{Scoville16}, scaled to dust temperatures of 100 K in the nuclei, after subtracting the potential synchrotron contribution.  The CO flux is converted to gas mass using $\alpha_\text{CO(2-1)} = 1.83\pm0.97$ M$_{\sun}$ (K km s$^{-1}$ pc$^{2}$)$^{-1}$, calibrated for NGC~6240 by \citet{cicone18}.}
\tablenotetext{c}{Estimated synchrotron contribution to continuum fluxes, based on 5 GHz emission from \citet{Beswick01} scaled by $\nu^{-1.06}$, the spectral index measured from VLA 8.4 GHz \citep{Carral90} and 15 GHz \citep{Colbert94} observations.}
\tablenotetext{d}{Corrected M$_{\text{BH}}$ is calculated by subtracting M$_{\text{H}_2}$ from M$_\text{central}$.}
\end{table*}

\subsection{Accretion Rates and Depletion Timescales}

\citet{Puccetti16} used \textit{Chandra} and \textit{NuSTAR} X-ray spectroscopy to model the accretion rates of the two black holes, and found intrinsic bolometric luminosities of 2.6 $\times10^{44}$ erg s$^{-1}$ and 8 $\times10^{44}$ erg s$^{-1}$ for the north and south nuclei, respectively.  Using L$=\eta$ $\dot{\text{M}}$ c$^{2}$ and assuming an energy efficiency $\eta$ of 10\%, we convert to a mass accretion rate of 0.05 and 0.14 M$_{\sun}$ yr$^{-1}$ for the north and south nuclei.  Assuming no replenishment, if all of the measured gas mass is accreted onto the black holes at this rate, the depletion timescales are therefore roughly 2 Gyr for the north and 5 Gyr for the south.  These timescales are significantly longer than a typical merger timescale, suggesting either that the accretion rate will increase or that some gas mass will be ejected (or both), unless considerable circumnuclear gas remains after the system is relaxed.  Indeed, post-starburst or early-type galaxies can still have nuclear reservoirs, but because they are not as likely to host an AGN as ongoing gas-rich mergers, it is plausible to imagine their nuclear reservoirs have been substantially reduced.

Averaging between continuum and CO(2-1) measurements, the gas mass in the south nucleus is approximately five times higher than that in the north.  The bolometric luminosities and therefore mass accretion rates differ by a factor of about three in the same sense.  A larger sample of black holes will show how closely the central molecular gas mass tracks the mass accretion rate of the black hole.

\subsection{Caveats to the Black Hole Mass Measurements}

This Letter in general assumes that the dynamical mass measurements of \citep{Medling15_bh} are correct and looks for additional contributions to the central point masses.  However, here we consider briefly some caveats associated with the original dynamical mass measurements, which use both Keplerian disk models and Jeans Axisymmetric Mass models \citep[JAM;][]{Cappellari08}.  The disk models assume entirely coplanar circular orbits; the JAM models include velocity anisotropy but are still axisymmetric.  Although \citet{medling11,nucleardisks,Medling15_bh} show evidence that the stellar dynamics within the sphere of influence of the black hole are relatively relaxed and disky, the systems are overall highly unvirialized and could contain complex orbits unaccounted for in the original modeling.  However, we stress that if the young stars are in a relaxed disk, the most stable dynamical state for the gas is too.  

The models of \citet{Medling15_bh} attempt to account for all (resolved) smoothly varying contributions to the mass profile.  That is, rather than assuming a (fixed or varying) mass-to-light ratio from the stellar light, they match the kinematics to a simple power-law mass profile to fit both the scaling factor and the index.  Although the stellar light from their imaging varies smoothly, our high-resolution CO(2-1) imaging shows that the CO is neither axisymmetric nor necessarily smoothly varying (although the continuum emission plausibly is).  If the CO is a more accurate tracer of the molecular gas and represents a dynamically important component to the mass profile, the assumptions of \citet{Medling15_bh} might break down.  However, we also note that the mass profiles measured by \citet{Medling15_bh} are dominated by the central point sources (the rest of the galaxy within the fitting region, $\sim$100pc, making up roughly 10\% of the mass).  Thus, errors in the details of the external gas or stellar distributions likely do not strongly affect the measured central point masses.


\section{Conclusions}
\label{conclusions}

We use long-baseline ALMA CO(2-1) and continuum measurements of the two nuclei of NGC~6240 to map the molecular gas within $\sim$40 pc of each black hole.  Our two independent molecular gas mass measurement techniques identify a substantial mass of gas below the spatial resolution limit of the 
original dynamical M$_\text{BH}$ measurements that could not be distinguished from the black holes.  In the south nucleus, and in the sum of the two, these corrections are sufficient to reduce the implied black hole masses to within the scatter of black hole scaling relations.

Our data confirm that molecular gas can play a substantial role in fueling AGN on tens of parsec scales, and reveal that dynamical black hole mass measurements must resolve this small scale -- or correct for the gas mass present --  to measure accurate black hole masses.  The two black holes in this work show different levels of correction, with gas masses making up 5\%-11\% of the original black hole mass measurement in the north and 6\%-89\% in the south black hole.  Future long-baseline ALMA data of dynamically measured black hole masses will indicate which level of correction is more typical.  The amount of gas near a quiescent black hole could be minimal compared to that around a gas-rich obscured AGN like NGC~6240; this variability must be characterized before statistical corrections can be made to other black hole mass measurements.  

The measured gas masses differ by a factor of five between NGC~6240's two nuclei, a larger difference than the ratio of their bolometric luminosities (approximately three).  If the gas mass in the central few tens of parsecs is instrumental in fueling the AGN (on short timescales), it may correlate well with the mass accretion rate of the black hole -- similar observations of a larger sample will test this prediction.  A substantial mass of gas within a black hole sphere of influence is likely to form a viscous accretion disk, which has important implications for subgrid prescriptions of black hole accretion rates in hydrodynamical simulations and for the timescales associated with accretion and feedback: a Bondi-Hoyle type accretion prescription will substantially overestimate the accretion rates.

\acknowledgments

The authors wish to recognize and acknowledge the very significant cultural role and reverence that the summit of Maunakea has always had within the indigenous Hawai'ian community; we are privileged to be guests on your sacred mountain.  
We also wish to pay respect to the Atacame\~{n}o community of the Chajnantor Plateau, whose traditional home now also includes the ALMA observatory.

Support for AMM is provided by NASA through Hubble Fellowship grant \#HST-HF2-51377 awarded by the Space Telescope Science Institute, which is operated by the Association of Universities for Research in Astronomy, Inc., for NASA, under contract NAS5-26555.  
GCP acknowledges support from the University of Florida.
ET acknowledges support from FONDECYT Regular 1160999 and 1190818, CONICYT PIA ACT172033, and Basal-CATA AFB170002 grants.
CC acknowledges funding from the European Union's Horizon 2020 research and innovation programme under the Marie Sk{\l}odowska-Curie grant agreement No. 664931.
CEM acknowledges support from the National Science Foundation under award numbers AST-0908796 and AST-1412851.

This paper makes use of the following ALMA data: ADS/JAO.ALMA\#2015.1.00370.S and \\
ADS/JAO.ALMA\#2015.1.00003.S. ALMA is a partnership of ESO (representing its member states), NSF (USA) and NINS (Japan), together with NRC (Canada) and NSC and ASIAA (Taiwan) and KASI (Republic of Korea), in cooperation with the Republic of Chile. The Joint ALMA Observatory is operated by ESO, AUI/NRAO and NAOJ.  The National Radio Astronomy Observatory is a facility of the National Science Foundation operated under cooperative agreement by Associated Universities, Inc.
Some of the data presented herein were obtained at the W. M. Keck Observatory, which is operated as a scientific partnership among the California Institute of Technology, the University of California and the National Aeronautics and Space Administration. The Observatory was made possible by the generous financial support of the W. M. Keck Foundation.  We enthusiastically thank the staff of the W. M. Keck Observatory and its AO team for their dedication and hard work.  

The analysis presented herein was initiated at the Aspen Center for Physics, which is supported by National Science Foundation grant PHY-1607611.
AMM, GCP, LBM, ET, and NS also thank the Sexten Center for Astrophysics, where the bulk of this paper was written.

%

\vspace{5mm}
\facilities{Keck:II (Laser Guide Star Adaptive Optics, OSIRIS), ALMA}

\bibliography{bibGOALS,bibNGC6240,bibBHs}
\bibliographystyle{aasjournal}



\end{document}